\documentstyle[preprint,prd,floats,aps]{revtex}

\begin{document}
\tighten

\title{
{\rm \hfill Preprint JINR E2-98-214 \vspace{1.5cm}}\\
Towards $N=2$ SUSY homogeneous quantum cosmology;
Einstein--Yang--Mills systems}
\author{Evgeni E. Donets\footnote{Email address: edonets@sunhe.jinr.ru}}
\address{Laboratory of High Energies, JINR, 141980 Dubna, Russia }

\author{Mikhael N. Tentyukov\footnote{Email address: tentukov@thsun1.jinr.ru} 
and Mirian M. Tsulaia\footnote{Permanent address: 
\hspace*{1em}Institute of Physics, Tamarashvili 6, Tbilisi 380077,
Georgia. Email address: tsulaia@thsun1.jinr.ru }}
\address{Bogoliubov Laboratory of Theoretical Physics,
 JINR, 141980 Dubna, Russia}

\maketitle

\begin{abstract}
The application of $N=2$ supersymmetric quantum
mechanics for the quantization of homogeneous systems
coupled with gravity is discussed.
Starting with the superfield formulation of an $N=2$ SUSY sigma model,
Hermitian self-adjoint expressions for quantum
Hamiltonians and Lagrangians for any signature of a sigma-model
metric are obtained. This approach is then applied to
coupled SU(2) Einstein-Yang-Mills (EYM) systems in axially symmetric
$Bianchi$-type I, II, VIII, IX,  $Kantowski-Sachs$, and closed
$Friedmann-Robertson-Walker$ cosmological models. It is shown that all
these models admit the embedding into the $N=2$ SUSY sigma model with the
explicit expressions for superpotentials being direct sums of
gravitational and Yang-Mills (YM) parts. In addition, the YM parts of
superpotentials exactly coincide with
the corresponding Chern-Simons terms.
The spontaneous SUSY breaking caused by YM instantons in EYM
systems is discussed in a number of examples.

\end{abstract}

\newpage
\section{INTRODUCTION}

In order to quantize a pure bosonic system one can apply supersymmetry
as a mighty tool for dealing with the problems of a quantum theory
\cite{WW}-\cite{CF}. The quantization can be done in two ways.
The first one is to embed the system in a four-dimensional supersymmetric
field theory and then reduce it to one dimension \cite{ED} - \cite{CH},
\cite{Moniz}
or, alternatively, to consider the desired Lagrangian as a bosonic part
of a supersymmetric sigma model after dimensional reduction
\cite{GB} - \cite{BG}. These two approaches are not equivalent in
general and the results can be different. The second method, i.e.,
the method of supersymmetric quantum mechanics, seems more
convenient for our purposes and we shall follow it hereafter.

In spatially homogeneous cosmological models the only dynamical
variable is time $t$; other (spatial) coordinates can be integrated 
out from the action.
Therefore, one can simply consider the corresponding mechanical
system and then try to make a supersymmetric sigma-model extension.
The case of pure gravity and gravity with scalar fields was
investigated recently by Graham and Bene in the
framework of $N=2$ supersymmetric (SUSY) quantum mechanics. 
However, construction of the
quantum Hamiltonian, proposed there, tunned out to be Hermitian not
self-dual  for the case
of indefinite signature of the metric in minisuperspace.
 In this paper we use
another construction of the corresponding Hamiltonian, which,
in accordance with general lines of quantization, is
 Hermitian self-adjoint for any type of signature
of the metric in minisuperspace. The obtained
quantum states coincide with those found in \cite{GB} , \cite{BG} only in
null fermion and filled fermion sectors, while
in other fermion sectors they exist only if the manifold,
determined by the minisuperspace metric, has corresponding nontrivial
cohomologies.

 We apply  developed $N=2$ SUSY sigma model technique for the
quantization of SU(2) Einstein-Yang-Mills (EYM) system in
homogeneous axially symmetric $Bianchi$ type I, II, VIII, IX, $Kantowski-Sachs$
(KS), and closed $Friedman-Robertson-Walker$ (FRW) cosmological
models. Since the work by Bartnik and McKinnon \cite{BK}
where  an infinite set of regular particle-like
SU(2) non-Abelian EYM configurations was obtained, further interest in the
EYM system has been caused by the unexpected properties
of their classical solutions. In particular, it has been
shown that non-Abelian EYM black holes violate the naive
``no-hair'' conjecture in an external region \cite{bhs}, as well as
demonstrating rather unusual internal structure \cite{DGZ} - \cite{BLM}
with the generic space-time singularity being  an infinitely
oscillating, but not of a mixmaster, type. The metric in the 
space-time region under an event horizon of a spherically symmetric
black hole is equivalent to the homogeneous cosmological Kantowski-Sachs
metric and this correspondence allows us to apply the methods
developed in quantum cosmology for the study of black hole singularities.
Classical EYM solutions in different (Bianchi) cosmologies
have still not been investigated so far, except the axially symmetric
Bianchi type I model, where chaotic behavior of the metric,
inspired by chaos in YM equations of motion, has been observed
\cite{KU} - \cite{Barrow}. In all the classical EYM systems mentioned above,
the nonlinear nature of the source YM field produces nontrivial
space-time configurations mainly in strong field regions, i.e.,
near black hole or cosmological space-time singularities, where
a pure classical description
of space-time should be replaced by a quantum field theory
and our present work is one step towards this goal.

 We show that all considered EYM models, containing initially
purely bosonic (gravitational and YM) degrees of freedom, admit $N=2$
supersymmetrization in the framework of the $N=2$ SUSY sigma model.
The inclusion of non-Abelian gauge fields to pure gravitational systems
produces additional parts in superpotentials, which, as we shall see below,
are equal to Yang-Mills Chern-Simons terms. The connection between the
superpotential and the ``winding number" in some supersymmetric
Yang-Mills field theories and sigma models was discussed
earlier \cite{ED} - \cite{CH}. However, direct generalization to
the EYM supersymmetric sigma models is not straightforward, since the
expression for the space-time metric, which in turn determines the
form of the {\em Ansatz} for the Yang-Mills field, can be arbitrary.
Therefore the fact that the $N=2$ supersymmetric
sigma model based on axially symmetric homogeneous EYM systems
respects this result is quite nontrivial.

 The paper is organized as follows. In Sec. II we discuss formal
aspects of $N=2$  SUSY  sigma models, starting with the
superfield
approach. In Sec. III the desired embedding of EYM systems
into the $N=2$ SUSY sigma model is described and explicit expressions
for superpotentials are given. The quantization and SUSY breaking
by YM instantons are discussed in Sec. IV. 

\section{$N=2$ SUSY QUANTUM MECHANICS}

   Let us first recall some main features of $N=2$ supersymmetric
quantum  mechanics, developed mainly in \cite{WW} - \cite{CF}.
We shall follow the superfield approach, since it is
more geometrical, rather than the component one, and
the component form of the corresponding Lagrangian obtained 
is obviously invariant under the desired SUSY transformations.
Consider superspace, spanned by the
coordinates $(t,\theta, \bar \theta)$, where $t$ is time, while
$\theta$ and its conjugate $\bar \theta$ are nilpotent Grassman
variables. The $N=2$ supersymmetry transformations in superspace with
the complex odd parameter $\epsilon$ have the following form
\begin{eqnarray}
\delta t &=&i \epsilon \bar \theta + i \bar \epsilon \theta \nonumber \\
\delta \theta &=& \epsilon \quad  \delta \bar \theta = \bar \epsilon,
\end{eqnarray}
which are generated by the linear differential operators
\begin{equation}
 \Omega = \frac{\partial}{\partial \bar \theta} +
i  \theta \frac{\partial}{\partial t}
\quad \mbox{and} \quad
\bar \Omega = \frac{\partial}{\partial \theta} +
i \bar \theta \frac{\partial}{\partial t}.
\end{equation}
Now one can introduce the main object of the theory -- the real
vector superfield $\Phi^i$:
\begin{equation}  \label{SF}
\Phi^i = q^i + \bar \theta \xi^i - \theta \bar \xi^i +
\bar \theta \theta F^i,
\end{equation}
where $q^i$ stands for all bosonic degrees of freedom of the system,
$\xi^i$ and $\bar \xi^i$ are their fermionic superpartners, and
$F^i$ is an auxiliary bosonic field. Since the superfield $\Phi^i$
transforms under the supersymmetry transformations as
\begin{equation}  \label{TR}
\delta \Phi^i = ( \bar \epsilon \Omega +  \epsilon \bar \Omega ) \Phi^i,
\end{equation}
the most general supersymmetric Lagrangian can be obtained in terms of the
supercovariant derivatives
\begin{equation}
D = \frac{\partial}{\partial \theta} -
i \bar \theta \frac{\partial}{\partial t}
\quad \mbox{and} \quad
\bar D =  \frac{\partial}{\partial \bar \theta} -
i  \theta \frac{\partial}{\partial t},
\end{equation}
which  anticommute with $\Omega$ and $\bar \Omega$; the resulting Lagrangian
\begin{equation} \label{LG}
L = \int d\theta d \bar \theta(-\frac{1}{2}g_{ij}
(D\Phi^i)(\bar D\Phi^i) + W)
\end{equation}
is invariant under supersymmetry transformations (\ref{TR}) by
construction and it corresponds to the one-dimensional
$N=2$ supersymmetric sigma model, characterized by the metric
$g_{ij}$ $(i,j = 1,...,n)$ of the ``target" manifold $M(g_{ij})$ and the
superpotential $W$, both  being a functions of the superfield $\Phi^i$.

Note that the Lagrangian (\ref{LG}) is $self-adjoint$ for any
signature of the metric $g_{ij}$.  This fact is especially important for
considering homogeneous systems coupled with gravity, since in these
cases
the manifold $M$ described by the metric $g_{ij}$ is not Riemannian.

After integration over the Grassman variables and elimination
of an auxiliary field $F^i$, one gets a more familiar component form
of the Lagrangian:
\begin{eqnarray} \label{CO}
L&=&\frac{1}{2}g_{ij}(q) {\dot{q}}^i  {\dot{q}}^j
+ i g_{ij}(q) \bar \xi^i ( {\dot{\xi}}^j + \Gamma^j_{kl}{\dot{q}}^k
\xi^l)  \nonumber \\
&&+ \frac{1}{2}R_{ijkl}\bar \xi^i  \xi^j \bar \xi^k \xi^l
-\frac{1}{2}g^{ij}(q) \partial_i W \partial_j W -
\partial_i  \partial_j W \bar \xi^i  \xi^j,
\end{eqnarray}
where $R_{ijkl}$ and $\Gamma^i_{jk}$ are the Riemann curvature
and Christoffel connection, corresponding to the metric $g_{ij}$.
The supersymmetry transformations can be also written
in the component form
\begin{eqnarray} \label{CP}
\delta q^i&=&\bar \epsilon \xi^i - \epsilon \bar \xi^i, \nonumber \\
\delta \xi^i&=&\epsilon(-i{\dot{q}}^i + \Gamma^i_{jk} \bar \xi^j  \xi^k
- \partial^i W), \nonumber \\
\delta \bar \xi^i&=&\bar \epsilon
(i{\dot{q}}^i + \Gamma^i_{jk} \bar
\xi^j
\xi^k - \partial^i W),
\end{eqnarray}
which allow us to find the conserved supercharges using the standard Noether
theorem technique:
\begin{eqnarray} \label{SZ}
Q&=&\xi^i(g_{ij}{\dot{q}}^j + i\partial_i W), \\
\nonumber
\bar Q&=&\bar \xi^i(g_{ij}{\dot{q}}^j - i\partial_i W).
\end{eqnarray}

Following the general lines of quantization of the system with
bosonic and fermionic degrees of freedom \cite{RC},
we introduce the canonical Poisson brackets
\begin{equation}
\{ q^i , P_{q^j} \}=\delta^i_j, \quad \{ \xi^i , P_{\xi^j} \}=-\delta^i_j ,
\quad \{ \bar \xi^i , P_{\bar \xi^j} \}=-\delta^i_j,
\end{equation}
 where $P_{q^j},P_{\xi^j}$, and $P_{\bar \xi^j}$ are momenta, conjugate
to $q^i$, $\xi^i$, and $\bar \xi^i$.
 After finding their explicit form
\begin{equation}
P_{q^i}= g_{ij}{\dot{q}}^i + i\Gamma_{j;ik}\bar \xi^j \xi^k,
\end{equation}
\begin{equation} \label{CN}
P_{\xi^i}=-ig_{ij}\bar \xi^j, \quad P_{\bar \xi^i} = 0,
\end{equation}
one can conclude from Eqs. (\ref{CN}) that the system possesses the
second class fermionic constraints
\begin{equation}
\chi_{\xi^i} =  P_{\xi^i} + ig_{ij}\bar \xi^j
\quad \mbox{and} \quad \chi_{\bar \xi^i} = P_{\bar \xi^i},
\end{equation}
 since
\begin{equation}
\{ \chi_{\xi^i}, \chi_{\bar \xi^j} \} = -i g_{ij}.
\end{equation}
Therefore, the quantization has to be done using the
Dirac brackets, defined for any  two functions $V_a$ and $V_b$ as
\begin{equation} \label{DR}
\{V_a , V_b \}_D =
\{V_a , V_b \} - \{V_a , \chi_c \} \frac{1}{\{\chi_c , \chi_d \}}
\{\chi_d , V_b \} .
\end{equation}
Using Eq. (\ref{DR}), one can easily find nonvanishing Dirac brackets between
bosonic and fermionic degrees of freedom:
\begin{equation}
 \{ q^i , P_{q^j} \}_D=\delta^i_j,
\quad \{ \xi^i , \bar \xi^j \}_D=-ig^{ij}.
\end{equation}
Then, after replacing the Dirac brackets with a graded commutator
\begin{equation}
\{ , \}_D \rightarrow i[,]_{\pm},
\end{equation}
one obtains the following (anti)commutation relations:
\begin{equation} [ q^i , P_{q^j} ]_-=i\delta^i_j,
\quad [ \xi^i , \bar \xi^j ]_+=g^{ij} .
\end{equation}

To make a quantum expression for
 supercharges (\ref{SZ})
it is convenient to introduce the projected fermionic operators
$\bar \xi^a = e_{\mu}^a \bar \xi^{\mu}$ and $\xi^a = e_{\mu}^a  \xi^{\mu}$
where $e^i_a$ is inverse to the tetrad $e_i^a$ ($e_i^a e^i_b=\delta^a_b$),
related to the metric $g_{ij}$ of the ``target" manifold $M$ and to the
metric of its tangent space $\eta_{ab}$
in the usual way, $e^a_ie^b_j \eta_{ab} = g_{ij}$.

However, the explicit  form of the supercharges
depends on the choice of operator ordering and
therefore is ambiguous. We take it as in \cite{CH}:

\begin{eqnarray} \label{UP}
Q&=&\xi^a e_a^i(P_i + i\omega_{iab}\bar \xi^a \xi^b
+ i\partial_i W) \nonumber \\
\bar Q&=&\bar \xi^a e_a^i(P_i + i\omega_{iab}\bar \xi^a \xi^b
- i\partial_i W) ,
\end{eqnarray}
where  $\omega_{iab}$ is the corresponding spin connection.

In what follows, we shall consider systems subject
to the classical Hamiltonian constraint
\begin{equation} \label{CC}
H_0 = \frac{1}{2}g^{ij}P_iP_j
 + \frac{1}{2}g^{ij}(q) \partial_i W \partial_j W = 0 ,
\end{equation}
 which in the quantum case should be replaced by the condition on
the quantum state $|\rho\rangle$,
\begin{equation} \label{QC}
H|\rho\rangle=0,
\end{equation}
with the  Hamiltonian
\begin{equation}
H = \frac{1}{2} [Q, \bar Q]_+,
\end{equation}
giving $H_0$ in the classical limit, i.e., when all fermionic fields
are set equal to zero.

 The important point is that the operators (\ref{SZ}) are nilpotent
and mutually Hermitian adjoint with respect to the measure
$\sqrt{|-g|}d^nq$ and, therefore,  the energy operator $H$ is
self-adjoint for any signature of the metric $g_{ij}$.
Now the Lagrangian (\ref{CO}) is self-adjoint after the fashion of construction,
since we use real superfields and hence the complex Noether charges
and their quantum mechanical expressions are Hermitian adjoint to each
other.

Obviously, now one can consider two first order differential equations
on the wave function,
\begin{equation} \label{NS}
\bar Q|\rho\rangle=0, \quad \mbox{and} \quad  Q|\rho\rangle=0,
\end{equation}
and therefore linearize the operator equation (\ref{QC});
the existence of normalizable solutions of the system (\ref{NS})
means, in turn, that supersymmetry is unbroken quantum mechanically.

In order to solve the system consider the Fock space spanned by the
fermionic creation and annihilation operators $\bar\xi^{a}$ and $\xi^{a}$,
respectively, with $[\xi^{a}, \bar\xi^{b}]_+ = \eta^{ab}$.
The general state in this Fock space is obtained in terms of
the series expansion
\begin{eqnarray}
|\rho \rangle &=&F(q)|0\rangle + ... +
\frac{1}{n!}
\bar\xi^{a_1}...\bar\xi^{a_n}F_{a_1...a_n}(q) |0\rangle  \nonumber \\
&&= F(q)|0\rangle + ... +
\frac{1}{n!}\bar\xi^{i_1}...\bar\xi^{i_n}F_{i_1...i_n}(q) |0\rangle,
\end{eqnarray}
 where the coefficients in expansions of this
series are $p$-forms defined on the manifold $M(g_{ij})$, and their
number due to the nilpotency of fermionic creation operators
is finite.
Since the fermion number operator $N= \bar\xi^{a}\xi_{a}$ commutes with
the Hamiltonian $H$ and
\begin{equation}
[N,Q]_- = - Q, \quad [N,\bar Q]_- = \bar Q,
\end{equation}
one can consider states characterized by the different fermion
numbers separately. Now the solution in empty and filled fermion
sectors is simply expressed in terms of the superpotential $W$ as follows:
\begin{equation} \label{EF}
|\rho_0\rangle = \mbox{const}\times e^{-W}|0\rangle,
\end{equation}
\begin{equation} \label{FF}
|\rho_n\rangle = \mbox{const}\times \frac{1}{n!}\bar\xi^{a_1}...\bar\xi^{a_n}
\epsilon_{a_1...a_n}
e ^{+W}|0\rangle.
\end{equation}

In order to investigate the solutions in other fermion
sectors, let us first recall \cite{ED} that in the case of vanishing
superpotential
operators $\bar Q_0$ and $Q_0$ (supercharges with $W=0$)
act on the $p$-forms $F$ as exterior and co-exterior derivatives,
respectively. So solution of the equation $\bar Q_0|\rho\rangle=0$
cannot be written as
\begin{equation} \label{BP}
|\rho_p\rangle=\bar Q_0|\sigma_{p-1}\rangle
\end{equation}
only if the corresponding $p$-th cohomology group $H^p(M)$
of the manifold $M(g_{ij})$ is
nontrivial.
Before generalizing this result to
 systems with nonzero superpotential $W$, first note that
\begin{equation} \label{SV}
\bar Q = e^{-W} \bar Q_0 e^{W} \quad \mbox{and} \quad
 Q = e^{W} Q_0 e^{-W}.
\end{equation}
Now, using Eqs. (\ref{BP}) and (\ref{SV}) one can prove
that the general solution in $p$-fermion sectors $(p =1,...,n-1)$
of the first equation in Eqs. (\ref{NS}) for the case of trivial
cohomology group $H^p(M)$ is
\begin{equation}
|\rho_p\rangle = \bar Q |\sigma_{p-1}\rangle.
\end{equation}
However, because $Q$ and $\bar Q$ are Hermitian adjoint
to each other, the second equation in Eqs. (\ref{NS})
indicates that this state has zero norm and consequently is unphysical.
Therefore the possible existence of supersymmetric ground states,
i.e., solutions of the zero-energy Shr\"odinger-type equation (\ref{QC}),
is directly related to the topology of the considered manifold
$M(g_{ij})$, since all states except those in purely bosonic
and filled fermion sectors can be excluded even without solving
the system (\ref{NS}), if the topology of the manifold $M(g_{ij})$
is trivial.

For purely bosonic systems with nonvanishing potential
energy the described $N=2$ supersymmetrization turns out to be
the simplest possible one and it can be applied for canonical
quantization of any appropriate homogeneous cosmological model
coupled with matter. After the choice of operator ordering in the
supercharges, Eq. (\ref{QC}) in the null fermion sector corresponds to
the Wheeler-DeWitt equation for the considered Einstein-matter
system and its solution (\ref{EF}) is then easily obtained in
terms of superpotential $W$, since SUSY allows us to linearize the
quantum Hamiltonian equation. \\

\section{$N=2$ SUPERSYMMETRIZATION OF SU(2) EINSTEIN-YANG-MILLS
COSMOLOGICAL MODELS}

Now we are in a position to make the $N=2$  supersymmetric extension
of homogeneous axially symmetric  SU(2) Einstein-Yang-Mills systems
given by the action
\begin{equation}
 S = \int  d^4x\sqrt{- G}( R - \frac{1}{2} F_{\mu \nu}^A
F^{A \mu \nu}).
\end{equation}

We restrict ourselves to a subclass of homogeneous
space-times which admit a representation in the form of an unconstrained
Hamiltonian system for a corresponding classical coupled system of
equations; i.e., we consider axially symmetric Bianchi type I, II,
VIII, IX (axially symmetric
Bianchi type VII is equivalent to Bianchi type I), Kantowski-Sachs  and
closed Friedmann-Robertson-Walker  cosmological models.

The general diagonal Bianchi type axially symmetric space-times are
parametrized by two independent functions of a cosmological time
$b_1(t)$ and $b_3(t)$,
\begin{equation} \label{ME}
ds^2 = - dt^2 + b^2_1(t)[(\omega^1)^2 + (\omega^2)^2] +
b^2_3(t)(\omega^3)^2, \
\end{equation}
where $\omega^i$ are basis left-invariant one-forms
($d\omega^i=\frac{1}{2}C^i_{jk} \omega^j \wedge \omega^k$) for the
 spatially homogeneous three-metrics, depending on three spatial (not necessarily
Cartesian) coordinates $x,y,z$:\\
Bianchi type I:
\begin{equation}
\omega^1 = dx,  \quad \omega^2 = dy, \quad \omega^3 = dz.
\end{equation}
Bianchi type II:
\begin{equation}
\omega^1 = dz,  \quad \omega^2 = dx, \quad
\omega ^3 = dy - xdz.
\end{equation}
Bianchi type VIII:
\begin{eqnarray}
\omega^1&=&dx + (1 + x^2) dy + (x - y - x^2y) dz, \nonumber \\
\omega^2&=&dx + (-1 + x^2) dy + (x + y - x^2y) dz, \nonumber \\
\omega^3&=&2x dy + (1 - 2xy) dz,
\end{eqnarray}
Bianchi type IX:
\begin{eqnarray} \label{BI}
\omega^1&=&\sin{z} dx - \cos{z} \sin{x} dy, \nonumber \\
\omega^2&=&\cos{z} dx + \sin{z} \sin{x} dy, \nonumber \\
\omega^3&=&\cos{x}dy + dz.
\end{eqnarray}

 As  was shown by Darian and Kunzle
\cite{KU}, the general ansatz for an SU(2) Yang-Mills
field, compatible with the symmetries of axially symmetric  Bianchi-type
cosmological models, is also expressed in terms of two independent
real-valued
functions $\alpha(t)$ and $\gamma(t)$ of a cosmological time only  and
has the form
\begin{equation} \label{AN}
A=\alpha(t)(\omega^1 \tau_1 + \omega^2 \tau_2) + \gamma(t)
\omega^3 \tau_3,
\end{equation}
where  $\tau_i$ are SU(2) group generators, normalized as
$[\tau_i,\tau_j]=\epsilon_{ijk}\tau_k$.

 Kantowski-Sachs space-time
\begin{equation}
ds^2 = - dt^2 + b^2_3(t) dr^2 + b^2_3(t) d\theta^2 +
b^2_1(t) (\sin \theta)^2 d\phi^2
\end{equation}
does not belong to Bianchi classification and admits an additional
spherical symmetry;
so SU(2) YM {\em Ansatz} has a different form, originating from the Witten
{\em Ansatz} for the static spherically symmetric case after the mutual replacement
$r \rightarrow t$, $t \rightarrow r$:
\begin{equation}
A_0= 0, \hspace{0.1cm} A_r=\gamma(t)L_1, \hspace{0.1cm}
A_{\theta}=-L_3 + \alpha(t)L_2, \hspace{0.1cm}
A_{\phi}=\sin\theta[L_2 + \alpha(t)L_3],
\end{equation}
\noindent
where $L_1=(\sin\theta \cos\phi, \sin\theta \sin\phi, \cos\theta)$,
$L_2=(\cos\theta \cos\phi, \cos\theta \sin\phi, -\sin\theta)$,
\newline
$L_3=(-\sin\phi, \cos\phi, 0)$ are spherical projections of SU(2)
generators.

 We also consider the closed Friedmann-Robertson-Walker model
separately, because its general YM {\em Ansatz} \cite{HO}
[SU(2) YM field on $S^3$] is not obtained from Bianchi type IX
after setting $\alpha(t)=\gamma(t)$ in Eq. (\ref{AN}).
The closed FRW model with the interval
\begin{equation}
ds^2 = -  dt^2 +  b^2(t) (d\chi^2 + \sin^2\chi
(d\theta^2 +\sin^2\theta d\phi^2))
\end{equation}
($\chi$,  $\theta$, and $\phi$ are angles on
$S^3$) admits the following representation for the SU(2) YM field,
expressed in terms of a single real-valued function $\alpha(t)$:
\begin{eqnarray}
A_0 = 0, \quad
A_j = \frac{1}{2}(\alpha(t) +1) U {\partial}_j U^{- 1} , \\
U = \mbox{exp} \{ i \chi[\sin \theta(\sigma_1\cos\phi + \sigma_2\sin\phi) +
\sigma_3\cos\theta]\}, \hspace{0.7cm}  j=1,2,3, \\ \nonumber
\end{eqnarray}
where $\sigma_i$ are Pauli matrices.

Inserting these {\em Ans\"atze} into the action and integrating over
all variables except $t$ one obtains the one-dimensional Lagrangian
\begin{equation}
L_0=\frac{1}{2}g_{ij}(q) {\dot{q}}^i  {\dot{q}}^j - V(q) = K - V; \
\end{equation}
 here, $g_{ij}(q)$ is the metric in the extended
minisuperspace i.e., in the configuration space of spatially
homogeneous axially symmetric three-metrics coupled with the
corresponding SU(2) Yang-Mills fields.

Let us consider the functions
$q^i = (b_1, b_3, \alpha, \gamma)$
as a bosonic components of the superfield  (\ref{SF}).
One can introduce the same number of fermionic fields
($\bar \xi^i$ and $\xi^i$) and therefore make $N=2$ supersymmetrization
of the Lagrangian $L_0$ if, and only if, the potential
$V(q)$ admits the expression via a function $W(q)$, called a superpotential:
\begin{equation} \label{WX}
V(q)=\frac{1}{2}g^{ij}(q)\frac{\partial W(q)}{\partial q^i}
\frac{\partial W(q)}{\partial q^j}.
\end{equation}
In this case $N=2$ SUSY Lagrangian (\ref{CO})
 and the corresponding Hamiltonian, obtained after usual Legendre
transformation, are self-adjoint for any signature of the metric
$g_{ij}$ in the extended minisuperspace.

The kinetic terms for all Bianchi and
Kantowksi-Sachs models are the same,
\begin{equation}
K= - {\dot{b_1}}^2 b_3 - 2 \dot{b_1}\dot{b_3}b_1 + {\dot{\alpha}}^2 b_3 +
{\dot{\gamma^2}}\frac{b_1^2}{2b_3},
\end{equation}
and the only difference between them is due to the potential terms.
Using the expression for the metric on the extended ``minisuperspace,''
\begin{equation} \label{BB}
g_{b_1b_1}=-2b_3, \quad g_{b_1b_3}=-2b_1, \quad
g_{\alpha \alpha}=2b_3, \quad g_{\gamma \gamma}=\frac{b_1^2}{b_3},
\end{equation}
and the explicit form of the potentials, we have found some
superpotentials as a solutions of Eq. (\ref{WX}), hence making an
 $N=2$ SUSY extension of the given Einstein-Yang-Mills systems.
The results are collected in the Table I.

\begin{table}[htb]
\caption{\label{thetable} Potentials and superpotentials}
\begin{tabular}{|c|c|c|}
  &   &  \\
  & Lagrangian $L_0$  & Superpotential \\
  &                     & $W=W_{gr}+{\bf W_{YM}}$ \\
  &   &  \\ \hline
     &   &       \\ 
{\scriptsize Bianchi }I & $K - [{\bf \frac{1}{b_3} \alpha^2 \gamma^2 +
\frac{b_3}{2 b^2_1} \alpha^4}];$ & $0 + {\bf \alpha^2\gamma};$ \\
     &   &   \\ \hline
     &   &       \\
{\scriptsize Bianchi }II & $K -
[{\bf \frac{1}{4}\frac{b_3^3}{b_1^2} }+ {\bf
\frac{1}{b_3} \alpha^2 \gamma^2 +
\frac{1}{2}\frac{b_3}{b^2_1} {(\alpha^2+\gamma)}^2}];$ &
$\frac{1}{2}b_3^2 + {\bf (\alpha^2\gamma + \frac{1}{2}\gamma^2)};$ \\
     &   &   \\ \hline
     &   &         \\
{\scriptsize Bianchi}  VIII & $K - [{\bf \frac{1}{4}\frac{b_3^3}{b_1^2} } + {\bf b_3} +
{\bf \frac{1}{b_3} \alpha^2 {\gamma}^2 +
\frac{1}{2}\frac{b_3}{b^2_1} {(\alpha^2-\gamma)}^2}];$ &
$\frac{1}{2}(2b_1^2 - b_3^2) +
{\bf (\alpha^2\gamma - \frac{1}{2}\gamma^2)};$ \\
    &    &   \\ \hline
    &   &          \\
{\scriptsize Bianchi } IX & $K - [{\bf \frac{1}{4}\frac{b_3^3}{b_1^2} } - {\bf b_3} +
{\bf \frac{1}{b_3} \alpha^2 {(\gamma-1)}^2 +
\frac{1}{2}\frac{b_3}{b^2_1} {(\alpha^2-\gamma)}^2}];$ &
$\frac{1}{2}(2b_1^2 + b_3^2) +
{\bf (\alpha^2(\gamma-1) - \frac{1}{2}\gamma^2)};$  \\
  &  & $or$ \\
  &   & $\frac{1}{2}(b_3^2 - 4b_1b_3) +
{\bf (\alpha^2(\gamma-1) - \frac{1}{2}\gamma^2)};$ \\
  &   &   \\ \hline
  &   &   \\
KS  & $K - [\frac{1}{b_3} \alpha^2 \gamma^2 - {\bf b_3} +
{\bf \frac{1}{2}\frac{b_3}{b^2_1} {(\alpha^2-1)}^2}];$ &
 $2b_1b_3 + {\bf \gamma(\alpha^2-1)};$ \\
   &   &  \\ \hline
   &   &    \\
FRW &
$-\frac{3}{2} b {\dot{b}}^2 + \frac{1}{2} b {\dot{\alpha}}^2 +
\frac{3}{2}b  - \frac{1}{2} {\bf \frac {(1 - \alpha ^2)^2}{b}};$ &
$ \frac{3}{2} b^2 +
{\bf (\frac{1}{3}\alpha ^3 - \alpha)};$ \\
  &    &
\end{tabular}\end{table}

One should note that the obtained superpotentials $W$ in all these cases
turn out to be direct sums of pure gravitational $W_{gr}$
(first listed in \cite{BG}  in terms of Misner variables) and Yang-Mills
parts $W_{YM}$. This fact is quite interesting and does not follow
{\it a priori}
from general expectations, since in the sigma-model approach
considered above gravitational and Yang-Mills variables in the
Lagrangian $L_0$ are not separated. Moreover, it seems that
the YM field is a unique one, which,  being
coupled with gravity, can allow the
corresponding superpotential to be in the form of a direct sum.
It follows from the following statement that the superpotential is also
the least Euclidean action -- solution of the Euclidean Hamilton-Jacobi
equation of the considered system. One can reconstruct from
the superpotential
the corresponding Euclidean solutions, those which give the main
contribution to the wave function in a quasiclassical approach.
So the gravitational part of the superpotential $W_{gr}$ determines
the Euclidean gravitational background configurations which
should not be changed if a matter field is added. It is possible
only if matter configurations do not contribute to the
energy-momentum tensor. The
Yang-Mills part of the superpotential $W_{YM}$ just provides such
a possibility since it produces self-dual YM instantons
with the energy-momentum tensor identically vanished. We discuss
this point in more detail in the next section.

Note that the full superpotential $W=W_{gr}+W_{YM}$ does not exist as
a solution of Eq. (\ref{WX}) if we cancel
one of the relevant YM function $\alpha$ or $\gamma$; there are
no nontrivial self-dual solutions of YM equations of motion with one of
YM functions canceled and  $W_{YM}$
ceases to exist in this case. The question about other solutions
of Eq. (\ref{WX}) which are not direct sums of
gravitational and YM parts is still open; however, it seems
unlikely that such solutions can be obtained in a closed
analytical form.

On the other hand, one more crucial observation can be done, that
for all considered models the Yang-Mills part of the superpotential
coincides with the corresponding Chern-Simons functional, calculated on a
three-dimensional slice $t=\mbox{const}$. Indeed, it can be checked that
the YM Chern-Simons terms
\begin{equation} \label{SN}
W_{YM}=\frac{1}{2}\int d^3x\sqrt{|-G|}\epsilon^{0 \lambda \mu \nu}
(A^a_{\lambda} \partial_{\mu}A^a_{\nu}
 + \frac{1}{3}f^{abc}A^a_{\lambda}A^b_{\mu}A^c_{\nu})
\end{equation}
 turn out to be solutions of the Euclidean Hamilton-Jacobi equation
and therefore play the role of the Yang-Mills part of the superpotential.
Such a coincidence of YM Chern-Simons terms (\ref{SN})
with YM superpotentials
(\ref{WX}) in the framework of the one-dimensional sigma model
describing a YM field coupled with gravity seems to be very
surprising. Definitely, this statement   is not
true in the general case of an
arbitrary space-time  and takes place for the suggested models
as a consequence of the symmetries of the space-time metrics and
corresponding YM {\em Ans\"atze}. Note, that there exist no  similar
expressions for the $W_{gr}$ part of the superpotential in terms of
a functional of gravitational variables except  the Bianchi type IX
model with a nonzero cosmological
constant,  where the Chern-Simons functional in terms of Ashtekar's
variables \cite{Ashtekar} is also  an exact solution of the
Ashtekar-Hamilton-Jacobi equation \cite{Kodama}.

So we have shown that the considered homogeneous axially symmetric
EYM systems admit an $N=2$ supersymmetric sigma-model extension
with the superpotentials given explicitly in  Table I and this
gives us a suitable background for the quantization. \\

\section{THE QUANTIZATION AND $SUSY$ BREAKING BY YM INSTANTONS}

\subsection{Supersymmetry at classical and quantum levels}
As can be seen from the supersymmetry
transformations (\ref{CP}),
in order to prevent $N=2$ SUSY breaking at the classical
level, the classical pure bosonic configurations must satisfy
the properties
\begin{equation}
{\dot q}^i(t) = 0 \quad \mbox{and} \quad \partial^iW\Bigl(q^i(t)\Bigr) =0,
\end{equation}
along with the classical Hamiltonian constraint (\ref{CC}).
Such classical configurations really exist in an usual
field theory in a flat space-time, and the simplest
well-known example is a scalar rest particle $(\dot {q^i} = 0)$
on a bottom of a potential with $V(q^i) = 0$.

In contrast with such examples, dealing with unconstrained
homogeneous systems with gravity included, any nontrivial
classical solution of Einstein (or Einstein coupled with a
matter) equations never has all momenta vanished,
${\dot q}^i(t) \not= 0$.
These systems satisfy Eq. (\ref{CC})
due to the dynamical balance between the kinetic and potential terms
with both positive and negative signs.

Hence, any homogeneous Einstein (or Einstein-matter) system,
being embedded into the $N=2$ supersymmetric sigma model never has
solutions of equations of motion with unbroken supersymmetry;
i.e., supersymmetry is always spontaneously broken at the ``tree level.''

Let us see what happens in the quantum mechanical approach.
In the Einstein-Yang-Mills systems considered above
the number of bosonic functions $q^i$ is 4, which is also
the fermion number of the filled fermion sector.
Therefore we shall consider solutions of the zero energy
Schr\"odinger-type equation (\ref{QC}) in these empty and
filled fermion sectors.

The superpotential $W(q)$ is always defined up to the sign,
since it is the ``square root" of the bosonic potential $V(q)$.
Both signs are physically acceptable and correspond to the solutions
in empty, Eq.  (\ref{EF}), and filled, Eq. (\ref{FF}), fermion sectors
when finding the supersymmetric wave functions.
The normalizability of bosonic wave function
for ``positive" superpotential  means in turn
the normalizability of filled fermionic wave function
for the ``negative" superpotential and vice versa.
We define the norm of the physical state as
$\pm \int \sqrt{|- g|} \langle \rho||\rho \rangle d^4q$
in order to avoid the problem  of the negative norm in
the four-fermion sector, caused by the timelike component of the
fermionic field. The plus sign in the definition of the
norm corresponds to $+W(q)$ while the minus sign has
to be taken as $-W(q)$.

 Let us accept for definiteness the positive sign of the
superpotential.
First consider pure gravitational systems, when $\alpha$
and $\gamma$ functions along with their fermionic partners
are set equal to zero. As was stated above, supersymmetry
is spontaneously broken for any nontrivial solutions
of Einstein equations. Quantum mechanically the supersymmetry
is restored for Bianchi type I, II, $\mbox{IX}_{(1)}$, Kantowski-Sachs and FRW
models since the solution of Eq. (\ref{QC}), $|\rho_0^{gr}\rangle
=\mbox{const}\times e^{-W_{gr}}|0 \rangle$, in the null
fermion sector is normalizable:
\begin{equation} \label{NO}
\int_0^{+\infty}db_1 \int_0^{+\infty}db_3
\sqrt{|-g|} e^{-2W_{gr}} < \infty.
\end{equation}

Therefore we are facing an interesting situation,  where
unlike ordinary supersymmetric quantum mechanics,
the supersymmetry being spontaneously broken at the ``tree level"
is then restored quantum mechanically.

The only exceptions are the second (in   Table I)
superpotential for Bianchi type $\mbox{IX}_{(2)}$
and Bianchi type VIII  where the supersymmetry remains
broken at the quantum level as well, since their norm (\ref{NO})
diverges at the upper limit.

Further inclusion of the Yang-Mills field spontaneously breaks
the supersymmetry again, because, as one can see from  Table I,
the Yang-Mills part of the superpotential
$W_{YM}$ for all considered models, being the corresponding Chern-Simons
term, is odd function of $\alpha$ and $\gamma$; consequently,
the YM parts of the wave function
$|\rho_0^{YM} \rangle=\mbox{const}\times e^{\pm W_{YM}}|0\rangle$
both in null and filled fermion sectors are not normalizable:
\begin{equation}
\int_{-\infty}^{+\infty} d\alpha \int_{-\infty}^{+\infty} d\gamma
\sqrt{|-g|} e^{\pm 2W_{YM}} \rightarrow \infty .
\end{equation}

In order to find possible supersymmetric wave functions in one-,
two-, and three-fermion sectors, one has to investigate the
topology of the extended minisuperspace. The simplest way of
doing that is going to the Misner parametrization \cite{MI}
of the space-time metric (\ref{ME}):
\begin{equation}
ds^2 = -N^2(t)dt^2 + \frac{1}{6}e^{2A(t) + 2B(t)}
[(\omega^1)^2 + (\omega^2)^2] +
\frac{1}{6}e^{2A(t) - 4B(t)}(\omega^3)^2 .
\end{equation}
In terms of Misner variables the metric in the extended
minisuperspace (\ref{BB}) has the simple diagonal form
\begin{equation}
g_{AA} = -1, \quad g_{BB} = 1, \quad
g_{\alpha \alpha} = 2 e^{-2A - 2B}, \quad
g_{\gamma \gamma} = e^{-2A + 4B},
\end{equation}
which shows that the topology of the extended minisuperspace
is equivalent to the Minkowski one with all cohomologies trivial,
$H^p(M(g_{ij}))=0,\quad p=1,2,3$, and, in accordance with the discussion of
Sec. II, no physical states in one-, two-, and three-fermion
sectors exist since they have zero norm. Similarly, there are no
physical states except the ones in null and filled fermion sectors in
the considered pure gravitational systems.  \\

\subsection{ A role of instantons}
Let us discuss in more detail the mechanism of spontaneous
supersymmetry breaking in the null fermion sector
when the YM field is added to a pure gravitational system
(such as Bianchi type I, II, $\mbox{IX}_{(1)}$, KS and FRW)
which is quantum mechanically supersymmetric,  since
it admits a normalizable zero-energy solution of the Wheeler-DeWitt
equation (\ref{QC}). This mechanism turns out to be quite similar to
the one considered in \cite{WW}, \cite{SH} - \cite{KM}
where the SUSY breaking by instanton configurations
has been discussed.

Indeed, as was already mentioned, the superpotential $W(q)$
(if exists) is one of the solutions of the Euclidean
Hamilton-Jacobi equation and represents a ``least" Euclidean action
of field configurations, giving the main quasiclassical contribution
into the wave function and providing
 the SUSY breaking
after  inclusion of the Yang-Mills field. The explicit form of the superpotential
allows us to reconstruct such classical configurations by solving the
first order system:
\begin{equation} \label{SD}
g_{ij}{\dot q}^j=-\frac{\partial (W_{gr} + W_{YM})}{\partial q^i} .
\end{equation}
For pure gravitational degrees of freedom these equations are
 equivalent to the (anti-)self-duality gravitational equations
$R_{\mu \nu \lambda \sigma}=\pm \tilde R_{\mu \nu \lambda \sigma}$
while  $W_{YM}(q)$ part of the superpotential in Eq. (\ref{SD}) gives rise
to the (anti-)self-dual  Yang-Mills equations
$F^a_{\mu \nu}=\pm \tilde F^a_{\mu \nu}$ on a given
gravitational background determined by the $W_{gr}$.

Then, (anti-)self-dual Yang-Mills instantons in our systems
can be interpreted as a tunneling solutions (with the nonvanishing
Euclidean action) between topologically distinct vacua. In this case
the YM instanton contribution provides the SUSY breakdown
due to the energy shift from the initial zero to some positive level and
this fact is expressed in the nonnormalizability of the YM part of the 
zero energy  wave function
$|\rho^{YM}_0\rangle=\mbox{const}\times e^{-W_{YM}}|0\rangle$.

As an illustration of these statements, let us consider Euclidean
configurations in Bianchi type IX and Kantowski-Sachs EYM systems.

 {\em Bianchi type IX system.} The solutions of Hamilton-Jacobi equation (\ref{SD}),
which correspond to the gravitational part of both possible superpotentials
$W_{gr(BIX_{(1)})}=\frac{1}{2}(2b_1^2+b_3^2)$ and
$W_{gr(BIX_{(2)})}=\frac{1}{2}(b_3^2 - 4b_1b_3)$,
have been discussed by Gibbons and Pope \cite{GP}. For our purposes we
would like to mention some of them  using a slightly  
different notation.

 One of the solutions of Eq. (\ref{SD}) with the normalizable
superpotential $W_{gr(BIX_{(1)})}$ turns out to be
the (anti-)self-dual Eguchi-Hanson \cite{EH} metric which has the form
\begin{equation}
ds^2=f^2dr^2 +
\frac{r^2}{4}[(\omega^1)^2 + (\omega^2)^2] +
\frac{r^2}{4}f^{-2}(\omega^3)^2,
\end{equation}
with
\begin{equation}
f^2 = {\Bigl[1 - {(\frac{a}{r})}^4\Bigr]}^{-1},
\end{equation}
and $\omega^{i}$ is determined by Eq. (\ref{BI}).
In order to bring this metric to
the form (\ref{ME}), one should introduce the ``Euclidean time" $r$ as
\begin{equation}
dt = {\Bigl[1 - {(\frac{a}{r})}^4\Bigr]}^{-\frac{1}{2}}dr .
\end{equation}
The Eguchi-Hanson metric has vanishing Euclidean action $S^{gr}_{EH}=0$, which
is completely determined by its surface contribution \cite{GH},
since the volume contribution is canceled identically ($R=0$ ``on shell")
for EYM systems.

 Inserting the expression for the metric functions
into the Hamilton-Jacobi equations for the Yang-Mills part
of the superpotential $W_{YM(BIX)}=-\alpha^2(\gamma-1)+\frac{1}{2}\gamma^2$
and differentiating with respect to the introduced variable $r$
one obtains the system
\begin{equation}
\dot{\alpha} = \frac{2}{r}f^2(\alpha\gamma-\alpha),
\end{equation}
\begin{equation}
\dot{\gamma} = \frac{2}{r}(\alpha^2-\gamma),
\end{equation}
which are the self-duality YM equations on an Eguchi-Hanson background
solved by the family of instanton solutions \cite{BC}
\begin{equation}
\alpha=\frac{a_1\sinh(\rho)}{\sinh[a_1(\rho+a_2)]}, \quad
\gamma=a_1\tanh(\rho) \coth[a_1(\rho+a_2)],  \quad
\frac{r^2}{a^2} = \coth(\rho),
\end{equation}
with the action $S^{YM}_{EH}=8\pi^2 (a_1^2-1)/2$ for $a_1>1$,
$a_2=0$, and $S^{YM}_{EH}=8\pi^2 (a_1^2)/2$ for $a_1>1$,
$0 < a_2 < \infty$,
where $a_1$ and $a_2$ are the
constants of integration.

  The extremal Euclidean configurations,
produced by the non-normalizable superpotential $W_{gr(BIX_{(2)})}$,
are self-dual Taub-NUT (Newman-Unti-Tamburino) gravitational instantons with nonvanishing
action \cite{TU}; similarly, the YM part
of the superpotential gives rise to the self-dual YM
instantons \cite{BC1} on a Taub-NUT background.

{\em The KS system.} For the EYM system in Kantowski-Sachs space-time
with $W_{gr(KS)}=2b_1b_3$
the gravitational degrees of freedom $b_1$ and $b_3$ obey
 the following self-duality equations:
\begin{equation}
\dot b_1 b_3 + \dot b_3 b_1 = b_3,
\end{equation}
\begin{equation}
\dot b_1 = 1,
\end{equation}
satisfied by
\begin{equation}
b_1 =t \quad \mbox{and} \quad b_3=1,
\end{equation}
which is nothing more than the flat Euclidean $R^4$ space-time metric
with $r$ and $t$ interchanged.
From the Yang-Mills part of Eq. (\ref{SD}) with
$W_{YM(KS)}=-\gamma(\alpha^2-1)$
one obtains
the usual YM (anti-)self-duality equations in $R^4$, written in the
``polar" coordinates
\begin{equation}
\dot{\alpha}  =  \alpha \gamma,
\end{equation}
\begin{equation}
\dot{\gamma}t^2= \alpha^2 - 1,
\end{equation}
with the well-known family of YM instanton solutions, having the 
topological charge $k=1$ \cite{PL}
\begin{equation}
\gamma = \dot{\psi} \quad \mbox{and} \quad \alpha = e^{\psi} \dot{g},
\end{equation}
 where
\begin{equation}
\psi = - \ln \left(\frac{1 - g^2}{2t}\right) , \quad g = \left(\frac{a_1 -t}{a_1 + t}\right)
\left(\frac{a_2 -t}{a_2 + t}\right).
\end{equation}

Note that the dimension of  moduli space $\mathcal M $ of SU(2)
Yang-Mills instantons with a topological charge $k$ on a given Riemannian  
$4-D$ manifold
${\bf \overline{M}}$ [which has
first
Betti number $c_1$
and the dimension $c_2^-$ of the maximal submanifold in cohomologies
$H^2({\bf \overline{M}}, R)$
where the corresponding intersection form is negatively defined]
is \cite{AH}
\begin{equation} \label{DI}
\mbox{dim}({\mathcal M}_{SU(2)}) = 8 k - 3(1 - c_1 + c_2^-),
\end{equation}
and in the simplest Kantowski-Sachs case with ${\bf \overline{M}}$ = $R^4$
($k=1$, $c_1=c_2^-=0$) is equal to 5. In the framework of
our approach, since we  quantize  the system reduced
to one dimension, only some of these instantons
are taken into account. In fact, we deal with the subclass of
all possible YM instantons, originating from the chosen {\em Ans\"atze},
which share the space-time symmetries in the Lorentzian sector.
However,  their contribution breaks the supersymmetry fatally
in conformity with the general expectations, as should take place
in a  full $4-D$ quantum theory.

 To summarize, it is shown that the spontaneous supersymmetry
breaking which takes place if the Yang-Mills field is added
to pure gravity is caused in a quasiclassical approach by
a YM instanton contribution to the wave function.
This contribution,
in accordance with general expectations, provides the energy
shift $\Delta E$ from a zero level. To estimate this energy
shift for EYM systems an instanton calculation technique
can be used, which also should give  the possibility to find the
lowest level normalizable wave function
$|\rho^1\rangle, $ $H |\rho^{EYM}_1\rangle = \Delta E |\rho^{EYM}_1\rangle$
for the considered models.
This work is in a progress now. \\

\section{CONCLUSIONS}

We would like to conclude with the following remarks.
The $N=2$ SUSY quantum mechanical sigma model approach allows us to
obtain conserved supercharges as being
Hermitian adjoint to each other, along  with the self-adjoint
expressions for the Hamiltonian and Lagrangian for any signature
of a sigma model metric. This gives the possibility to use the
supersymmetry as a tool for the quantization of various
homogeneous systems coupled with gravity
if they can be embedded
into the considered $N=2$ SUSY sigma model.
The desired embedding has been done for coupled SU(2)
EYM systems in some cosmological models which admit explicit
expressions for the superpotentials as being direct sum of
gravitational and Yang-Mills parts. After the quantization
the only nontrivial zero-energy wave functions in null and
filled fermion sectors  turns out to have a diverging
norm and this fact indicates spontaneous breaking of supersymmetry,
caused by YM instantons.

\acknowledgments
We would like to thank E. A. Ivanov, G. Jackeli, S. O. Krivonos,
A. P. Nersessian, M. S. Volkov, and especially A. I. Pashnev for
helpful discussions and comments. This work was supported in
part by Russian Foundation for Basic Research Grant 96-02-18126.
The work of M. M. T. was also supported in part by Grant INTAS-96-0308.

\end{document}